\documentclass[12pt]{article}

\usepackage{amssymb,amsmath}

 \makeatletter
\def\alt{\mathrel{\mathpalette\gl@align<}}
\def\agt{\mathrel{\mathpalette\gl@align>}}
\def\gl@align#1#2{\lower.6ex\vbox{\baselineskip\z@skip\lineskip\z@
\ialign{$\m@th#1\hfil##\hfil$\crcr#2\crcr\sim\crcr}}}
\makeatother

\begin{document}
\begin{flushright}
{\tt hep-ph/0703107}\\
MIFP-07-07 \\
March, 2007
\end{flushright}
\vspace*{1.0cm}

\begin{center}
\baselineskip 20pt
{\Large\bf
%Unity of Gauge, Higgs and Matter Fields \\
%and their Interactions from an Orbifold GUT
Yukawa Couplings in a Model with \\ Gauge, Higgs and Matter Unification
}

\vspace{1cm}

{\large
Ilia Gogoladze$^a$, Chin-Aik Lee$^a$, Yukihiro Mimura$^b$ and Qaisar Shafi$^a$}
\vspace{.5cm}

{\baselineskip 20pt
\it
$^a$Bartol Research Institute, Department of Physics \& Astronomy, \\
University of Delaware, Newark, DE 19716, USA \\
\vspace{2mm}
$^b$Department of Physics, Texas A\&M University,
College Station, TX 77843-4242, USA
}
\vspace{.5cm}

\vspace{1.5cm}
{\bf Abstract}
\end{center}

We discuss how unification of the gauge, Higgs and  (three chiral
family) matter superfields can be realized from the compactification
of a six dimensional supersymmetric $SU(8)$ gauge theory  over the
orbifold $\mathbb{R}^4 \times T^2/\mathbb{Z}_3$. The bulk gauge
interaction includes Yukawa interactions to generate masses for
quarks and leptons after the electroweak symmetry is broken. The
Yukawa matrices in this case turn out to be antisymmetric, and thus
not phenomenologically viable. To overcome this we introduce
 brane fields which are vector-like under the standard model gauge symmetry,
 and so do not alter the number of chiral families. In such a setup, the
observed fermion masses and mixings can be realized by taking into
account suppression effects from the effective Wilson line couplings
and large volume of the extra dimensions.

\thispagestyle{empty}

%\bigskip
\newpage

\addtocounter{page}{-1}

\section{Introduction}
\baselineskip 18pt

The discovery of atmospheric and solar neutrino oscillations have
yielded convincing evidence for new physics beyond the standard model (SM).
Many possible extensions of the SM have been proposed, and
one of the most compelling ideas is quark-lepton unification
and that of grand unified theories (GUTs) \cite{Pati:1974yy,Georgi:1974sy}.
Combined with supersymmetry (SUSY), one obtains a rather attractive framework
for physics beyond the standard model.
%
%Supplemented with some
%additional, typically flavor symmetries,
%
Gauge couplings can be unified in the minimal SUSY standard model (MSSM),
and the masses and mixings of quarks and leptons can be also explored
in a (more or less) unified setup.
However, the Higgs sector of these theories is
usually problematic, leading to well known problems of fine-tuning,
doublet-triplet splitting, dimension five nucleon decay, etc.

More recent discussion of SUSY GUTs often invoke one or more extra
dimensions to circumvent some of these problems.
After dimensional reduction, a vector/tensor field in some higher
dimensional compactified space decomposes into  a number of scalar,
vector and tensor components in four dimensions (4D). The left- and
right-handed Weyl fermions in 4D are unified in higher dimensional
fermions. The idea of compactification has been applied to break
symmetries via orbifold boundary conditions
\cite{Kawamura:1999nj,Hall:2001pg,Mimura:2002te}. When applied to
GUT models, the colored Higgs particles can be projected out, while
the unprojected zero mode of the Higgs doublet remains light.
Dimension five
proton decay mediated by  the colored Higgs fields is forbidden in
the model~\cite{Kawamura:1999nj}. Though the gauge symmetry is
explicitly broken by the orbifold conditions, the gauge couplings
can still unify provided that the brane localized gauge interactions are
suppressed by the large volume of the extra dimension~\cite{Hall:2001pg}.
In such a framework, the idea of gauge-Higgs
unification~\cite{Manton:1979kb} was  recently revived
\cite{Dvali:2001qr}. The scalar Higgs fields can be unified with the
gauge fields in some higher dimensional vector field(s). Through
the orbifold boundary condition, the higher dimensional gauge
symmetry gets broken since the generators of the associated 4D gauge
bosons are projected out.  The broken generators for the extra
dimensional components can have massless modes, and the Wilson line
operator can be identified as the Higgs bosons breaking  the
symmetry that remains  in 4D \cite{Hosotani:1983xw}. More precisely, via a
Hosotani transformation, we can show that the resulting model is equivalent
to one with a trivial Wilson line and non-Abelian orbifold projections breaking
electroweak symmetry.
The SUSY non-renormalization theorem protects the scalar from acquiring a large
mass and which therefore survives at low energy.
Interestingly,
this idea is compatible with  the extension of the SM to large gauge symmetries
including GUTs.
%Interestingly,

An interesting consequence of gauge-Higgs unification is that
Yukawa interactions can arise from the gauge interaction when
fermions are also higher dimensional bulk
fields~\cite{Burdman:2002se,Gogoladze:2003bb}. The 4D zero modes of
fermions can be chiral due to  orbifold projections, and the higher
dimensional extension of the fermion kinetic term with covariant
derivative
can include Yukawa couplings, with some of the higher
dimensional components of the gauge fields identified as Higgs fields. In the
left-right symmetric realization of such a model~\cite{Shafi:2002ck},
the matter representation for achieving gauge-Yukawa unification
can be much simpler than that of the SM construction, and indeed,
unification of gauge and Yukawa coupling constants can be
realized~\cite{Gogoladze:2003bb}. In the models over a 5D ${\cal N}
=1$ SUSY $S^1/\mathbb{Z}_2$ orbifold with bulk gauge symmetries such
as $SO(11)$ and $SU(8)$, which break down to
 $SU(4)_c \times SU(2)_L \times SU(2)_R$ in 4D,
matter fields are unified in hypermultiplets, and
 all three gauge couplings and third generation
Yukawa couplings (top, bottom, tau and Dirac tau neutrino) can be
unified. The top quark mass as well as the MSSM parameter
$\tan\beta$, a ratio of vacuum expectation values
(VEVs) for up- and down-type Higgs, are predicted. The prediction of
the top quark mass is in good agreement with the experiment. Thus,
unification of the gauge and Yukawa coupling constants can be an
important signal of extra dimensions at ultra high energy scales
\cite{Gogoladze:2003pp}.

In SUSY extensions, the matter fields  can be unified in higher
dimensional gauge multiplets \cite{Watari:2002tf,Li:2003ee},
especially if the model consists of $N=4$ vector multiplet in 4D
language, which can arise from  6D ${\cal N}=(1,1)$ SUSY.
Interestingly, three chiral families  can be obtained in the case of
a $T^2/\mathbb{Z}_3$ orbifold \cite{Watari:2002tf}. The three
families originate from the three chiral supermultiplets in the
$N=4$ gauge multiplet. Since three is the maximal number of chiral multiplet in
4D, this may explain family replication.

A hypermultiplet in the  adjoint representation of the bulk gauge
symmetry in 5D ${\cal N}=1$ SUSY $S^1/\mathbb{Z}_2$ orbifold model
can be incorporated into the gauge multiplet in 6D ${\cal N}=(1,1)$
SUSY orbifold models. In Ref.\cite{Gogoladze:2003ci}, it is found
that all the matter fields  for one family, the Higgs doublets,  as
well as gauge fields of the  SM can be unified in 6D ${\cal N} =
(1,1)$ SUSY $SU(8)$ gauge multiplets with $T^2/\mathbb{Z}_6$
orbifold. The three gauge couplings and the third generation Yukawa
couplings can also be  unified in the model. Alternative gauge groups
\cite{Gogoladze:2003yw} and  extensions including seven dimensional
models \cite{Gogoladze:2005az,Gogoladze:2006ps} have been discussed.
Since no other bulk matter fields can be introduced, the model can
explain why only the third family is heavy. A shortcoming of this type of model
is that there is no unique way of specifying the discrete $\mathbb{Z}_6$ orbifold
projections so that we get the matter and Higgs fields as the zero modes.
The first and second families are treated as
brane
fields to cancel the  brane localized gauge anomalies. The Yukawa
couplings for the first and second families are suppressed by a
large volume factor, but there is no good  reason as to why  the
mass of the first family is hierarchically small. Thus,
the masses and mixings are introduced by hand. If the  gauge
symmetry is extended to a group such  the  $SO(16)$, the two
families  can be included in the vector multiplet
\cite{Gogoladze:2003yw}, though the discrete charge assignment is
more complicated.

If we choose a $T^2/\mathbb{Z}_3$ orbifold for the $SU(8)$ model,
on the other hand, the discrete charge assignment is simple and
almost unique if  $N=1$ SUSY survives  in  4D. In this case,
the three chiral  families, the Higgs fields as well as the gauge
fields are naturally unified in one multiplet. However, the Yukawa matrix is
antisymmetric, which stems  from the fact that
the chiral superfields in the gauge multiplet are in  adjoint
representations of  the bulk gauge symmetry. As a consequence, after
electroweak symmetry  breaking, two families have degenerate
masses, and the first family is massless. One may introduce  brane
localized interactions to break the mass degeneracy through
cancellation. This looks not only unnatural, but is also inconsistent
with the volume suppression of the brane interactions. Without the
latter, gauge coupling unification can be adversely affected by
brane localized couplings.

In this paper, we will construct a phenomenological viable
three-family model from a  $T^2/\mathbb{Z}_3$ orbifold construction
based on $SU(8)$, such that the three chiral families, as well as
the gauge and Higgs fields are all unified. We will introduce  brane
localized fields which are vector-like under  the SM group, and
which are needed to cancel the  gauge anomalies originating from the
additional $U(1)$ gauge symmetries.
 We will discuss how the
 second family masses  can be suppressed due to the large volume of the
extra dimensions. The mass of the first family can be further
suppressed by a mechanism involving Wilson line operators.

This paper is organized as follows: In section 2, we will construct
a 6D ${\cal N} = (1,1)$ SUSY $T^2/\mathbb{Z}_3$ orbifold model with
$SU(8)$ bulk gauge symmetry,  which breaks down to $N=1$ SUSY with
$SU(4)_c \times SU(2)_L \times SU(2)_R \times U(1)^2$ at the 4D
fixed points. The three chiral matter multiplets of the SM and the
Higgs fields are  obtained from  the bulk gauge multiplet.
The Yukawa and gauge interactions are unified.
%at $M_{GUT}$.
In section 3, we  study how the fermion mass hierarchy
can be realized.  Section 4 explores  fermion masses and mixings in
more detail. Our conclusions are summarized in  Section 5.

\section{Gauge, Higgs and Three-Family Unification}

In this section we will construct a model in which the 4D gauge,
Higgs superfield and three families of matter superfield are unified
in a
 6D ${\cal N}=(1,1)$ SUSY gauge (super)multiplet associated with
 the gauge group $G=SU(8)$.
The gauge multiplet consists of a vector field,  four real scalars,
and both left- and right-handed Weyl fermions. The maximal
$R$-symmetry in ${\cal N}= (1,1)$ SUSY is $Sp(2)_L \times Sp(2)_R$.
The vector field is a singlet under the $R$-symmetry and decomposes
into a 4D vector field $A_\mu$ and two real scalar fields $A_5$
and $A_6$. The four real scalars in the gauge multiplet transform as
$({\bf 2},{\bf 2})$ under the $R$-symmetry, while the
left(right)-handed Weyl spinors transform as $({\bf 1},{\bf 2})$
 $(({\bf 2},{\bf 1}))$. From a 4D point of view,
there are four left-handed Weyl spinors.
All together, these fields are reorganized in a single  $N=4$ gauge
multiplet, which consists of one $N=1$ vector superfield $V$ and
three chiral superfields $\Phi_i$ $(i=1,2,3)$ in 4D. The scalar
component of $\Phi_1$ is $A_5- i A_6$. [The formalism of  6D models
is described  in Ref.\cite{Marcus:1983wb}.]

The two extra dimensions are compactified over a flat
$T^2/\mathbb{Z}_3$ orbifold. The orbifold transformation $\bf R$ is
$z \to \omega z$, where $z= x_5 + i x_6$ and $\omega = e^{2\pi
i/3}$. The transformation $\bf R$ can also act on the internal symmetry of the
Lagrangian, which in our class of models is the product of
$Sp(2)_L$, $Sp(2)_R$ and $Aut(G)$. This extension of $\bf R$ can
break SUSY as well as the  bulk gauge group $G$. Depending on the
discrete charge assignment, the 4D $N=4$ SUSY can be broken down to
$N$ = 0, 1  or 2.

If at least $N=1$ SUSY survives at a 4D fixed point, the orbifold
conditions for the superfields $V$ and $\Phi_i$ are
\begin{eqnarray}
V(x^\mu,\bar\omega \bar z,\omega z) &=& U V(x^\mu,\bar z,z) U^{-1}, \\
\Phi_1(x^\mu,\bar\omega \bar z,\omega z) &=& \bar \omega \,U \Phi_1(x^\mu,\bar z,z) U^{-1}, \\
\Phi_2(x^\mu,\bar\omega \bar z,\omega z) &=& \bar \omega^l \,U \Phi_2(x^\mu,\bar z,z) U^{-1}, \\
\Phi_3(x^\mu,\bar\omega \bar z,\omega z) &=& \bar \omega^m \,U \Phi_3(x^\mu,\bar z,z) U^{-1},
\end{eqnarray}
where $U$ is an $SU(8)$ matrix. Since there are  higher dimensional
versions of the gauge interaction term ${\rm tr}\,\Phi_1 [\Phi_2,
\Phi_3]$ in the Lagrangian, the condition  $1+l+m =0$ (mod 3) needs
to be satisfied. The only possible choices are $(l,m) =
(1,1),(2,0),(0,2)$. For the case $(l,m)=(2,0)$, $\Phi_1$ and
$\Phi_2$ form a hypermultiplet, and $V$ and $\Phi_3$ form an $N=2$
vector multiplet. In the same way, $N=2$ SUSY remains for $(l,m) =
(0,2)$. Thus, $(l,m)=(1,1)$ is the unique choice to obtain $N=1$
SUSY at the 4D fixed points with the orbifold conditions on all
three chiral superfields being the same. We will choose $l=m=1$
hereafter.

The $SU(8)$ matrix $U$ is chosen to be
\begin{equation}
U = {\rm diag} (1,1,1,1,\bar \omega,\bar \omega,\omega,\omega),
\end{equation}
 resulting in the breaking of the $SU(8)$ gauge symmetry to
$G_{422} \times U(1)_A \times U(1)_B$ in 4D, where $G_{422}$ is the
gauge symmetry $ SU(4)_c \times SU(2)_L \times SU(2)_R$
\cite{Pati:1974yy}.
The $SU(8)$ adjoint representation $\bf 63$  decomposes as
\begin{equation}
{\bf 63} \to
 \left(
  \begin{array}{ccc}
     ({\bf 15},{\bf 1},{\bf 1})_{0,0} & ({\bf 4},{\bf 2},{\bf 1})_{1,-1} & ({\bf 4},{\bf 1},{\bf 2})_{1,1} \\
     (\bar{\bf 4},{\bf 2},{\bf 1})_{-1,1} & ({\bf 1},{\bf 3},{\bf 1})_{0,0} & ({\bf 1},{\bf 2},{\bf 2})_{0,2} \\
     (\bar{\bf 4},{\bf 1},{\bf 2})_{-1,-1} & ({\bf 1},{\bf 2},{\bf 2})_{0,-2} & ({\bf 1},{\bf 1},{\bf 3})_{0,0}
  \end{array}
 \right)
 \oplus ({\bf 1},{\bf 1},{\bf 1})_{0,0} \oplus ({\bf 1},{\bf 1},{\bf 1})_{0,0}\,,
\end{equation}
\sloppy
where the generators of $U(1)_A$ and $U(1)_B$ subgroups are chosen and normalized to
be
${\rm diag}\,(1,1,1,1,-1,-1,-1,-1)/2$ and ${\rm diag}\,(0,0,0,0,1,1,-1,-1)$,
respectively.
Although the representations
are vector-like under the subgroup,
the zero modes of the superfields $\Phi_i$ are chiral after orbifold
projection,
and thus a chiral field theory in 4D is obtained.
%
%From the zero modes of the chiral superfields $\Phi_i$,
%we have three copies of
The zero modes
%which are three copies of
%
correspond to three copies of
$({\bf 4},{\bf 2},{\bf 1})_{1,-1}$ and $(\bar{\bf 4},{\bf 1},{\bf 2})_{-1,-1}$
representations,
and also three copies of
%as well as
$({\bf 1},{\bf 2},{\bf 2})_{0,2}$\footnote{Note that the Higgs bidoublet is
also chiral with respect to one of the $U(1)$ symmetries.}.
%which are chiral under the 4D gauge symmetry.
These can be interpreted as the
$SU(2)_L$ ($SU(2)_R$) doublet matter chiral superfields
$\Psi_i$ ($\Psi_i^c$) for quarks and leptons,
and the Higgs bidoublets $H_i$, respectively. As a
result, the  gauge and Higgs fields as well as the  three families
of SM matter fields plus three right-handed neutrinos are unified
in one gauge multiplet.

\fussy

There is only one pair of Higgs doublets in the MSSM. Indeed, if more
than two pairs of doublets survive well  below
the unification scale,
%$M_{GUT}$,
they can
spoil gauge coupling unification. The brane interaction of $H_i$
with the $G_{422}$ singlets can make two of the bidoublets massive,
so that  only one linear combination remains light.

It is worth noting that obtaining the three chiral families does not
depend on the details of the discrete charge assignments. Three
copies of chiral fields are always obtained if $N=1$ SUSY remains
in 4D. The converse  of this statement also holds.

Due to the $SU(8)$ bulk gauge unification, the $SU(4)_c$, $SU(2)_L$
and $SU(2)_R$ gauge couplings from a 4D point of view are unified at
the cutoff $M_*$ \cite{Hall:2001pg,Dienes:1998vh},
\begin{equation}
g_c^2 = g_{2L}^2 = g_{2R}^2 = g^2 \equiv \frac{g_{6D}^2}{V} = \frac{\hat g_{6D}^2}{V M_*^2},
\end{equation}
where $V$ is the volume of the extra dimensions. The 6D gauge
coupling $g_{6D}$ is a dimensionful parameter,  which can be turned
into a dimensionless coupling $\hat g_{6D}$ by employing the cutoff
$M_*$. A brane localized gauge kinetic term can modify this
unification, but this can be suppressed if  $V M_*^2$ is sufficiently
large.

Since the scalar component of  $\Phi_1$ is a higher dimensional
gauge field, the bulk gauge interaction includes the term ${\rm
tr}\,\Phi_1 [\Phi_2, \Phi_3]$, which contains the Yukawa couplings,
\begin{equation}
g\, \epsilon_{ijk} \Psi_i \Psi_j^c H_k,
\end{equation}
with conventional normalization of the gauge coupling, ${\rm
tr}\, T^a T^b = 1/2\, \delta^{ab}$, the bulk Yukawa coupling
constant is the same as the gauge coupling $g$.

Because  the chiral superfields are in the  adjoint representation,
the bulk Yukawa coupling is antisymmetric. One family is naturally
predicted to be massless, which, to a good approximation, is
desirable as far as the first generation is concerned. However, the
two non-zero mass eigenvalues of the fermions are degenerate, which
is a terribly wrong prediction. Although a brane localized
interaction, $y_{ijk}^\prime \Psi_i \Psi_j^c H_k$, may be introduced
to solve the problem, such interactions are suppressed by the volume
factor. This suppression is needed, as previously mentioned, to
preserve  gauge coupling  as well as gauge-Yukawa unification. In
addition, the fermion mass hierarchy between the second and third
families would then arise from fine-tuning, which is not attractive.
In the next two sections we will show how phenomenologically viable
fermion masses and mixings can be achieved in this class of models.

\section{Fermion Mass Hierarchy}

We have  three chiral families
$\Psi_i \,({\bf 4},{\bf 2},{\bf 1})_{1,-1}$,
$\Psi^c_i\, (\bar{\bf 4},{\bf 1},{\bf 2})_{-1,-1}$
and $H_i\, ({\bf 1},{\bf 2},{\bf 2})_{0,2}$ from the zero modes.
There are gauge anomalies at each 4D fixed point \cite{Scrucca:2004jn}
involving the $U(1)$ symmetries. In order to cancel these anomalies and retain
three chiral families, we introduce brane fields which are vector-like under
$G_{422}$ but not with respect to the $U(1)$'s. To
break the $U(1)$ symmetries, some $G_{422}$ singlet fields with
suitable $U(1)$ charges are also introduced.

Suppose that brane superfields $\bar \Psi^c_b \, ({\bf 4},{\bf
1},{\bf 2})$, and $\Psi^c_b \, (\bar{\bf 4},{\bf 1},{\bf 2})$ are
introduced with the following interaction:
\begin{equation}
\int d^6 x\, \delta(x_5)\delta(x_6)\int d^2\theta\,
\left(M \bar\Psi^c_b (\Psi_b^c +r_i \Psi_i^c) + y_{ik} \Psi_i \Psi^c_b H_k\right).
\end{equation}
The mass scale $M$ may be related to the $U(1)_{A,B}$ breaking scale.
The fermion mass matrix (e.g. for up-type quark) is given by
\begin{equation}
(\begin{array}{cccc}
   u_1 & u_2 & u_3 & \bar U^c_b
 \end{array})
\left(
 \begin{array}{cccc}
   0 & a_3 & -a_2 & m_1 \\
   -a_3 & 0 & a_1 & m_2 \\
   a_2 & -a_1 & 0 & m_3 \\
   r_1 M & r_2 M  & r_3 M & M
 \end{array}
\right)
\left(
 \begin{array}{c}
   u^c_1 \\ u^c_2 \\ u^c_3 \\ U^c_b
 \end{array}
\right),
\label{mass-matrix}
\end{equation}
where $a_i = g \langle H_{iu}^0 \rangle$ and $m_i = y_{ik} \langle
H_{ku}^0 \rangle$. If there is no other up-type Higgs field with a
large VEV, we have $a \equiv {(a_1^2+a_2^2+a_3^2)}^{1/2} = g v_u$. We
neglect the brane localized coupling $y^\prime_{ijk} \Psi_i \Psi_j^c
H_k$. Due to  volume suppression, $y_{ik}$ is suppressed, and thus
$m_i \ll a$. For the same reason, the dimensionless coefficients
$r_i\gg 1$. Without loss of generality, the $a_2$, $a_3$
and $r_3$ entries can be eliminated by the transformations $\Psi_i^\prime =
V_{ij} \Psi_j$ and $\Psi_i^{c\prime} = V_{ij} \Psi_j^c$, where $V$
is a unitary matrix. Thus, we can write the up-type quark mass
matrix as
\begin{equation}
M_u = \left(
 \begin{array}{cccc}
   0 & 0 & 0 & m_1^\prime \\
   0 & 0 & a & m_2^\prime \\
   0 & -a & 0 & m_3^\prime \\
   r_1^\prime M & r_2^\prime M  & 0 & M
 \end{array}
\right).
\label{mass-matrix2}
\end{equation}
With $M \gg v_u$, the mass eigenvalues can be calculated to be
\begin{eqnarray}
&&m_u^2 m_c^2 m_t^2 = \frac{a^4 r_1^{\prime2} m_1^{\prime2}}{1+r^2}, \\
&&m_u^2 m_c^2 + m_u^2 m_t^2 + m_c^2 m_t^2 =
   a^2 \frac{(a+r_2^\prime m_3^\prime)^2 + r^2 m_1^{\prime2}
  + r_1^{\prime2} (a^2+m^2)}{1+r^2},\\
&&m_u^2 + m_c^2 + m_t^2 = a^2 +
\frac{(a+r_2^\prime m_3^\prime)^2 + r_2^{\prime2} (m_1^{\prime2}+m_2^{\prime2})
+ r_1^{\prime2} (a^2 + m^2)}{1+r^2},
%\frac{(a-qz)^2}{1+r^2} +
%(x^2+y^2)\frac{p^2+q^2}{1+p^2+q^2}+
%p^2 \frac{a^2+z^2}{1+p^2+q^2},
\end{eqnarray}
where $r^2= r_1^2+r_2^2+r_3^2$ and $m^2 = m_1^2+m_2^2+m_3^2$.
Therefore, under the assumption of volume suppression ($a \gg m$ and
$r \gg 1$), the quark masses are hierarchical if $r_1^\prime \ll r$
is satisfied and  one finds
\begin{equation}
m_t \simeq a, \quad
m_c \simeq \frac{a}{r} + m_3^\prime, \quad
m_u \simeq a \frac{r_1^\prime}{r}\frac{m_1^\prime}{m_c}\,.
\end{equation}
For this case, the top, bottom, tau (and also Dirac tau neutrino)
Yukawa couplings are unified at the GUT scale together with the gauge coupling:
\begin{equation}
g_c = g_{2L} = g_{2R} = y_t = y_b = y_\tau = y_{\nu_{\tau}}.
\end{equation}

The scalar components of  $\Psi_1$ and $\Psi_1^c$, as well as of
$H_1$, are identified with the transverse components $A_z$ of the gauge field.
Because $A_z$ can  always be gauged away on the branes, it is not
possible to introduce brane couplings to the zero modes of the first
family. In the limit when there is no brane interactions for
$\Psi_1$, $\Psi^c_1$ and $H_1$, we obtain $a_2,a_3,m_1 \to 0$ and $r_1 \to
0$. The first generation is then massless, while the second
and third family masses are hierarchical.

Since the first family is not truly massless  we need a small
correction to the  brane coupling. There could exist a gauge
invariant coupling involving a Wilson line closed path with a
nontrivial winding number around the torus. This coupling is expected to be
exponentially suppressed \cite{Csaki:2002ur,Ibanez:1986ka},
which could provide an explanation of the
observed suppression of the first family mass terms \cite{wilson}.

\section{Fermion Masses and Mixings}

In the previous section we have seen how a fermion mass hierarchy
can be realized through suppression of the Wilson line couplings and
from volume suppression. In this section we discuss the Yukawa
matrices in more detail. Since a left-right symmetry survives at 4D
fixed points, the quark CKM mixings  vanish if only one bidoublet is
light, which is assumed for  successful gauge coupling unification.
Though it is an  attractive feature at  leading order, we have to
break the proportional relationship between the up- and down-type
quark mass matrices, as well as lepton Dirac mass matrices, to
obtain  realistic quark and lepton masses and mixings
\cite{King:2005bj}.

In our setup, the breaking of $G_{422}$ down to the SM can happen
by the VEVs of brane fields, which can break the proportional
relationship of the fermion mass matrices.
For example,
suppose that there is a nonrenormalizable coupling,
$\Sigma \Psi^c_i \bar\Psi_b^c S/M_*$,
in addition to the coupling $\Psi^c_i \bar\Psi_b^c S$,
where $S$ is a singlet under $G_{422}$
with appropriate $U(1)$ charges to make
the coupling gauge-invariant
and $\Sigma$ is a triplet under $SU(2)_R$.
%
%the renormalizable coupling, $\Sigma \Psi^c_i \bar\Psi_b$, can make
%
Then, $r_i$ in the fermion mass matrix such as Eq.(\ref{mass-matrix})
can be slightly different for up- and down-type quarks.
%
%and lepton mass matrices,
%if $\Sigma$ transforms as $({\bf 1},{\bf 1},{\bf 3})$ and/or $({\bf
%15},{\bf 1},{\bf 1})$ with appropriate $U(1)$ charges to make
%the coupling gauge-invariant.
%
%Its VEV will then play the dual role
%of breaking some of $G_{422}$ and one of the $U(1)$ symmetries.
%
There is enough freedom to fit fermion masses
and mixings if we include this type of couplings
with $SU(4)_c$ and $SU(2)_R$ adjoint fields.

Consider one such modification with
 $(r_3/r_2)_u = 1+\delta/2$ and $(r_3/r_2)_d = 1-\delta/2$ for example.
Then, in the basis used in Eq.(\ref{mass-matrix2}),
the quark mass matrices can be
written as
\begin{equation}
M_u = \left(
 \begin{array}{cccc}
   0 & 0 & 0 & m_1^u \\
   0 & 0 & a_u & m_2^u \\
   0 & -a_u & 0 & m_3^u \\
   \epsilon_u r_u M & r_u M  & 0 & M
 \end{array}
\right), \quad
M_d = \left(
 \begin{array}{cccc}
   0 & 0 & 0 & m_1^d \\
   0 & 0 & a_d & m_2^d \\
   0 & -a_d & 0 & m_3^d \\
   \epsilon_d r_d M^\prime & r_d M^\prime  & \delta r_d M^\prime & M^\prime
 \end{array}
\right).
\end{equation}
Integrating out the heavy vector-like quark fields,
we obtain
\begin{equation}
M_u^{3\times3} = \left(
 \begin{array}{ccc}
   0 & m_1^u & 0 \\
   \epsilon_u a_u & \frac{a_u}{r_u} + m_3^u & 0  \\
   0 & m_2^u & a_u
 \end{array}
\right), \quad
M_d^{3\times3} = \left(
 \begin{array}{ccc}
   0 & m_1^d & 0 \\
   \epsilon_d a_d & \frac{a_d}{r_d} + m_3^d & \delta a_d  \\
   0 & m_2^d- \delta \frac{a_d}{r_d} & a_d
 \end{array}
\right).
\end{equation}
The bulk field for the right-handed second family is
switched into the bulk field $\Psi^c_b$ in the large $r_{u,d}$
limit. In the expression for $3\times3$ mass matrices, the second
and third rows are switched. It is easy to find that $V_{cb} \simeq
\delta$, and one can also fit the relations
 $V_{ub} \sim V_{us} V_{cb}$ and $V_{cb} \sim m_s/m_b$.
The (1,1) elements in the quark mass matrices are tiny since the
effective Wilson line coupling is suppressed. This is
useful to explain the empirical relation $V_{us} \sim
\sqrt{m_d/m_s}$.
The hierarchy within the up- and down-type quarks, e.g. $m_u/m_t
\ll m_d/m_b$, is not explained in this model.

For the neutrino sector, the nonrenormalizable coupling
$\Psi_i^c \Psi_j^c \bar\chi^c \bar\chi^c$
is needed to obtain heavy right-handed Majorana masses.
The brane field $\bar\chi^c$ transforms as $({\bf 4},{\bf 1},{\bf 2})$ under $G_{422}$.
The observed bilarge neutrino mixing is not automatically realized  in
the model, but there is enough freedom to fit the neutrino data. The
large (essentially maximal) atmospheric neutrino mixing may be
explained by the choice $r_2 \simeq r_3$.
The model displays antisymmetry to leading order under the exchange
of
$\Phi_2$ and $\Phi_3$, which is related to the $Sp(2)_R$ symmetry in the bulk.
As in the quark sector,
due to the left-right gauge symmetry,
the mixing angles from the neutrino
 Dirac sector are effectively  small in the basis
where the charged-lepton Yukawa matrix is diagonal. Namely, even if
$r_2 \simeq r_3$, a large mixing in the charged-lepton Yukawa matrix
is essentially canceled by the mixing in the Dirac neutrino Yukawa
matrix. When considering type I seesaw,
%
%the large mixings are canceled with the Dirac neutrino Yukawa coupling.
%Thus,
%
the observed large mixing needs to be generated by suitably choosing
the right-handed Majorana mass matrix, irrespective of the symmetry.
If we consider type II seesaw
\cite{Schechter:1980gr}, on the other hand,
there is no reason to cancel the large mixing in the charged-lepton
Yukawa matrix
and
the exchange symmetry between $\Phi_2$ and
$\Phi_3$ can be the origin of the large atmospheric neutrino mixing.

\section{Conclusion and Discussion}

We have studied a 6D ${\cal N} = (1,1)$ SUSY model with $SU(8)$ bulk
gauge symmetry compactified over a flat orbifold $T^2/\mathbb{Z}_3$.
The $SU(8)$ symmetry is broken through orbifold compactification
 to  $SU(4)_c \times SU(2)_L
\times SU(2)_R \times U(1)_A \times U(1)_B$. The three chiral
families of the SM together with three right-handed neutrinos, as
well as Higgs fields, are the zero modes. Thus, gauge, Higgs and
matter fields of the  SM are unified in a single gauge
supermultiplet  in higher dimension.
However, this leads an antisymmetric
Yukawa coupling matrix for the three chiral families. As a consequence,
two of the fermion mass eigenvalues are degenerate, which is not
acceptable. To solve this problem, one could consider the orbifold
$T^2/\mathbb{Z}_6$  \cite{Gogoladze:2003ci}, so that
 the undesired eigenstate is projected
out,  only one family arises from  the bulk fields, and the third
family Yukawa couplings are unified with the gauge couplings. In
this paper, on the other hand, we have constructed a realistic model
employing brane localized fields which form  vector-like pairs under
the  SM. Being  vector-like,  these fields do not alter the number of
chiral families. In this setup, the fermion mass hierarchy can be
realized as a result of suppression of the effective Wilson line
couplings as well as the large volume of the extra dimensions. The
third family Yukawa couplings maintain their asymptotic unification
with the gauge couplings.  Ignoring threshold effects at both the
unification and weak scales yields
 $m_t = 178$ GeV
and $\tan\beta = 51$ \cite{Gogoladze:2003pp,Chkareuli:1998wi}.

While  $SU(8)$  was employed  to construct our model,  we briefly
comment on other choices for the   bulk gauge group. We find that
except for $SU(8)$, a  complete set of three chiral quark and lepton
families as well as Higgs fields is hard to achieve  for a
$T^2/\mathbb{Z}_3$ orbifold. In some cases, the SM particles are
partially included in the bulk fields, while in some other cases, more than
three chiral families are included in the bulk fields and a flavor gauge
symmetry survives in  4D. In the latter case, it seems better to work
with the $T^2/\mathbb{Z}_6$ orbifold. If the bulk fields only
partially include the SM matter fields, the other remaining chiral
fields are brane localized fields. In such partially unified models,
only the top quark Yukawa coupling can be unified with the gauge
couplings, and this  may help  explain the up-down hierarchy such as
$m_u/m_t \ll m_d/m_b$.

%
%One of the interesting case for such partial unification scenario
%is $SO(12)$ bulk gauge symmetry which breaks down
%$SU(5)\times U(1) \times U(1)$.
%The three copies of matter and Higgs ${\bf 10}_{1,1}$, $\bar{\bf 5}_{2,-1}$ and $\bar{\bf 5}_{-3,0}$
%are included in the $SO(12)$ adjoint $\bf 66$ bulk gauge multiplet.
%%
%In the choice of flipped hypercharge assignment,
%only top quark Yukawa coupling (and tau neutrino Dirac Yukawa coupling) can unify with
%the gauge couplings.
%The right-handed charged-leptons and
%down-type Higgs fields needs to be brane fields in this choice.
%Since the light up-type Higgs doublet at weak scale,
%as well as the first family,
%will be the higher dimensional component of gauge fields $A_z$,
%the effective Wilson line coupling for up quark mass
%is strongly suppressed.
%In this setup of the model,
%the up-down hierarchy $m_u/m_t \ll m_d/m_b$ can be explained.

\section*{Acknowledgments}

This work is supported in part by the DOE Grant \# DE-FG02-91ER40626
(I.G.),  (C.L.) and (Q.S.).

\end{document}